\documentclass[%
preprint
linenumbers,
 amsmath,amssymb,
 aps, physrev,
]{revtex4-2}

\usepackage{graphicx}
\usepackage{dcolumn}
\usepackage{bm}

\begin{document}

\preprint{APS/PRE}

\title{Trade-Off Between Multiplicity and Specificity in the Inter-layer Connectivity of non-identical Multilayer Networks}

\author{Aradhana Singh}
\email{aradhana22singh@gmail.com}
\affiliation{Department of Physics, Indian Institute of Science Education and Research, Tirupati, Andhra Pradesh 517507, India
}

\author{Amod Rai}
\affiliation{Department of Physics, Indian Institute of Science Education and Research, Tirupati, Andhra Pradesh 517507, India}

\author{Sheksha Dudekula}
\affiliation{Department of Physics, Indian Institute of Science Education and Research, Tirupati, Andhra Pradesh 517507, India}

\author{Antonio Palacios}
\affiliation{Department of Mathematics, San Diego State University, San Diego, USA}

\author{Devanarayanan P}
\affiliation{Department of Physics, Indian Institute of Science Education and Research, Tirupati, Andhra Pradesh 517507, India}

\date{\today}

\begin{abstract}

We study the coupled dynamics of multilayer networks with symmetric (ML$\text{s}$) and asymmetric (ML$\text{as}$) inter-layer connections. The symmetric inter-layer connections arise from a one-to-one correspondence between the nodes of different layers. In contrast, asymmetry results from the multiplicity of inter-layer connections, achieved by randomizing the links while preserving their overall density, thereby allowing one-to-many inter-layer connections.
We investigate how different types of inter-layer coupling impact the dynamics of non-identical multilayer networks. We find that the specificity of one-to-one inter-layer connections facilitates intra-layer synchronization (ILS). In contrast, for networks with random inter-layer connectivity, ILS depends on how randomness affects intra-layer homomorphism (the set of permutations that preserve the network structure). Furthermore, amplitude death (AD) in ML$\text{s}$ is observed at lower connectivity strength and frequency mismatch than the ML$\text{as}$. Moreover, AD in  ML$\text{s}$  depends on the density and topology but does not depend on the size of the networks. On the other hand,  AD in ML$\text{as}$ is influenced by network size in addition to density, topology, and inter-layer mismatches.
Moreover, both the ML$\text{s}$ and ML$\text{as}$ exhibit multi-stability, with the faster layer exhibiting a remanent periodic phase-locked oscillation, irrespective of the topology and inter-layer connectivity. In addition, remnant synchrony between nodes with homomorphic relationships is observed in the slower layer. Overall, we propose that symmetric inter-layer connections should be preferable for achieving intra-layer synchronization—regardless of global synchronization—and for sustaining permanent memory in multilayer networks with mismatched nodes across layers. However, to mitigate AD at low coupling values and layer mismatch, asymmetric inter-layer connectivity is more advantageous.  
\end{abstract}
\pacs{05.45.Xt,05.45.Pq}
\maketitle

\noindent \section{Introduction}
\noindent How nature works has been a quest of humanity for a long time, and in solving this, Networks are serving as a powerful tool. 
The structure of a network influences its dynamics, including factors such as error and attack tolerance \cite{Robust}, disease spread \cite{P_V_Mieghem_Rev}, synchronizability \cite{syn_book_kurths, clustering_syn}, and controllability \cite{controllability}. Additionally, networks can experience suppression of oscillations shown by individual units and stabilization of a homogeneous steady state, a phenomenon known as amplitude death (AD) \cite{AD_strogatz}.

\begin{figure}
\begin{center}
\includegraphics[width=1\columnwidth]{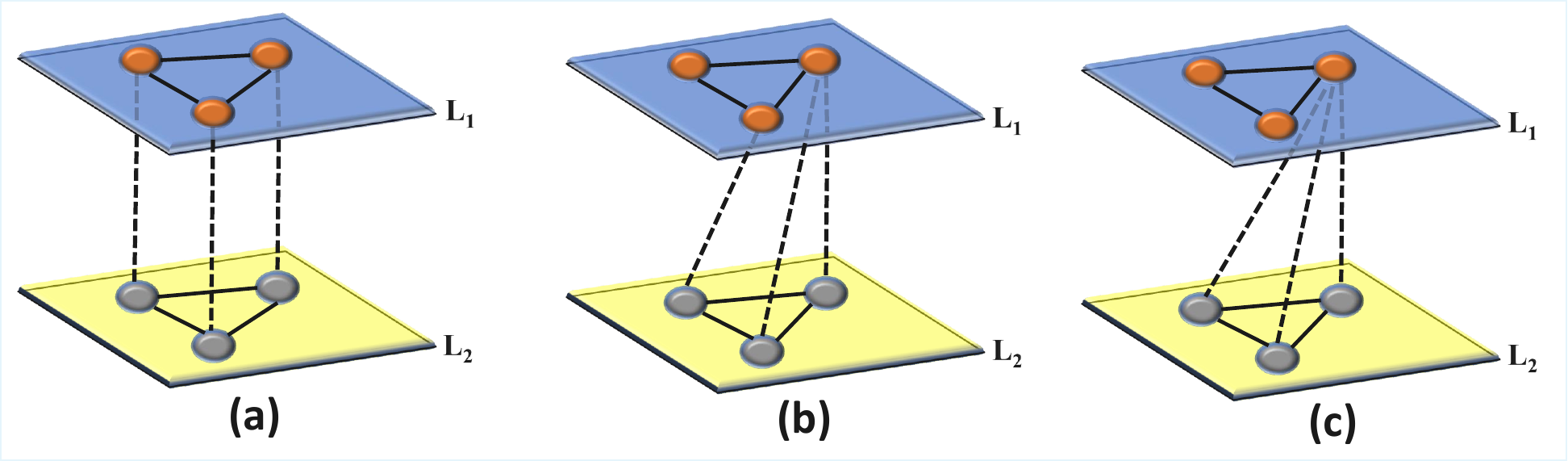}
\end{center}
\caption{{\bf Bi-layer small network with one-to-one and one-to-many inter-layer connections}. Displaying the simplest bi-layer network with different inter-layer configurations. Subplot (a) illustrates one-to-one inter-layer connectivity, while diagrams (b) and (c) depict the two possible one-to-many inter-layer connection configurations. The dotted (solid) lines represent inter-(intra-) layer connections. In our simulations, we use a similar strength of inter-layer and intra-layer connectivity to solely see the impact of the inter-layer connection topology.  
}
\label{Schematic-3nodes}
\end{figure}

\noindent Moreover, in the world of networks of networks, it is essential to adopt the multilayer network perspective to understand the overall dynamics. A multilayer network considers multiple channels of connectivity \cite{dedomenico}. This rich framework helps us understand many system properties that the single-layer network might not be able to explain. 
The multilayer perspective is crucial in nonlinear systems, where even a small perturbation can grow sufficiently to make the system unstable. The disturbance may arise from another layer, indicating a different type of connectivity among the nodes, thereby making the multilayer perspective even more essential. 
The following examples illustrate the existence of multiple layers of interactions in most real-world networks. In the brain, the different layers consist of neurons connected through synapses and gap junctions \cite{Multilayer-Brain}. In transport networks, the different layers can represent various modes of transportation \cite{transport_multilayer}. In the social network, the different layers can comprise people spread across various geographical scales or from different cultures \cite{boccaletti2014structure}. 
 The existing literature on the dynamics within a layer of a multilayer network has shown that these dynamics differ from those of an isolated network. A multilayer coupling has been shown to induce intra-layer synchronization, explosive synchronization, and disease localization among the layers \cite{syn_multilayer, explosive_syn, multi_layer_disease}. The activity of one layer is reported to be significantly affected by the structure and topology of other layers; for example, the activity of a layer can be suppressed or amplified by adding another layer \cite{VADAKKAN2025134715}.
Information transfer between nodes in a network has also been seen to have maximized by multiplexing the network with additional layers \cite{diffusion_dynamics}. The delay and different coupling in the layers promote the AD  \cite{AD-Delay, AD-different-coup}. The multilayer framework has also been utilized to study the robustness of networks against errors and attacks, considering various network topologies and error and attack strategies \cite{Robust_multi1, Robust_multi2}. Sparse multiplex networks, analogous to the multilayer networks with one-to-one connectivity, have been shown to exhibit better synchronization than sparse isolated networks \cite{Jalan_2016}, which becomes more pronounced when the layers are non-identical \cite{sync_layer_mismatch}.   

\begin{figure} [t]
\includegraphics[width=1\columnwidth]{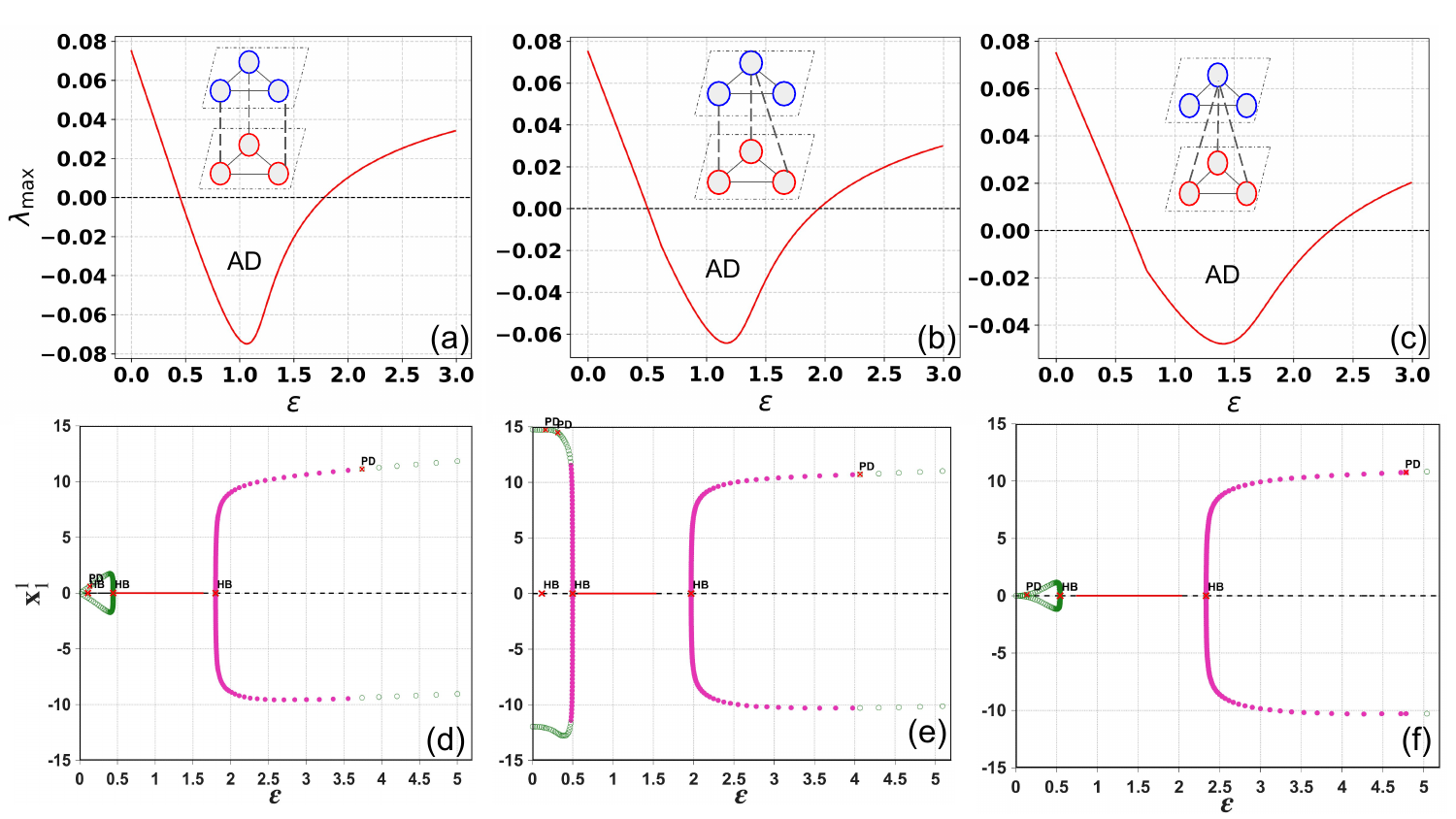}
\caption{{\bf Hopf bifurcation for the one-to-one and one-to-many inter-layer connectivity w.r.t. coupling strength ($\varepsilon$).} The plots (a-c) in the top panel illustrate how the largest eigenvalue of the Jacobian varies for the equilibrium state at the origin. The plots (d-f) in the bottom panel show the bifurcation plots for the three different inter-layer connectivity schemes displayed in the top panel. Here, the solid (dashed) lines display the stable (unstable) stationary states, and the closed (open) circles represent the stable (unstable) periodic states. Additionally, the various acronyms in the figure have the following meanings: HB: Hopf Bifurcation, PD: Period Doubling Bifurcation. In all the cases we have considered, $\omega_1=1$, $\omega_2=1.4$. 
}
\label{Bifurcation-3nodes}
\end{figure}

\noindent Furthermore, the connections between nodes of the same layer (Intra-layer connections) and between nodes of different layers (inter-layer connections)  are found to be crucial in understanding the dynamics of the entire network. The dynamics have been found to be connectivity-type specific, and an exchange between them has been shown to suppress the synchronization \cite{good_links_bad_links}. The resilience to network failures also depends on the interconnections and correlations across the layers, which have been reported to provide better resilience to random failures. Meanwhile, anti-correlated network layers are found to be robust against errors \cite{Robust_multi1}. 
Hence, it is worth wondering if we could tweak the inter-layer connectivity of a multilayer network to deliberately drive the network towards AD, Intra-layer synchrony, or desynchrony, or better yet, to devise a strategy for designing multilayer networks that minimize or maximize the required level of AD and Intra-layer synchronization. 

\begin{figure} [t]
\includegraphics[width=1\columnwidth]{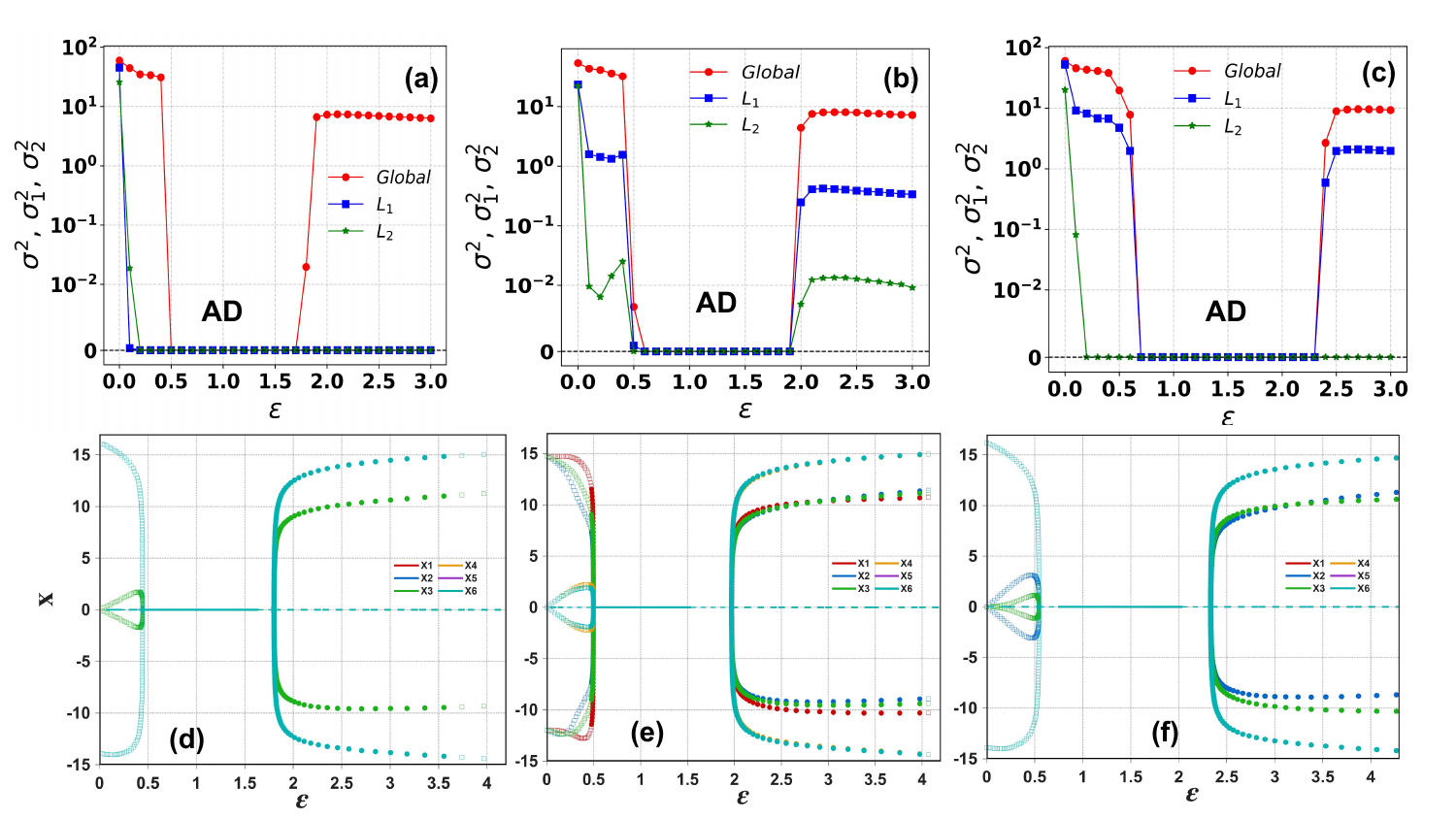}
\caption{{\bf Synchronization for the one-to-one and one-to-many inter-layer connectivity} The top panel shows the variation of the synchronization order parameters ($\sigma_1, \sigma_2$, and $\sigma$ for three different inter-layer configurations of $6$ nodes bi-layer networks. The bottom panels display the bifurcation plots for all nodes in these configurations, further illustrating the range of coupling for AD and synchronization. Subplots (a) and (d) show synchronization of all the nodes of the same layers in the non-AD regime. Subplots (b) and (e) show no intra-layer synchronization, whereas subplot (e) shows the cluster synchronization between the two nodes from the second layer, with a homomorphic relation unaffected by the inter-layer connectivity. Finally, subplots (c) and (f) exhibit intra-layer synchronization in the second layer, where all nodes maintain a homomorphic relationship with each other. In contrast, in the first layer, the two nodes synchronize, as they do not have inter-layer connections and so share a homomorphic relationship. 
For all these plots $N_{l_1} = N_{l_2} = 3$, and $\omega_1=1$, $\omega_2=1.4$ for all above cases.
}
\label{Synchronization-3nodes}
\end{figure}

\noindent Moreover, the inter-layer connectivity has been reported to affect the dynamics. Specifically, \cite{inter_intra_type} has shown that the type of interconnections plays a crucial role in the synchronous behavior. Additionally, \cite{Symmetry-cluster-syn} demonstrates that the symmetries of each layer and inter-connectivity between layers influence the synchronization patterns. Another study on multilayer networks demonstrated that asymmetrical inter-layer connection patterns lead to random multilayer networks with improved synchronizability \cite{Symmetries_syn-multilayer}. 
However, these studies are limited to studying synchronization and their stability with no focus on multi-stability, bifurcation, and AD. 

\begin{figure} [t]
\includegraphics[width=0.8\columnwidth]{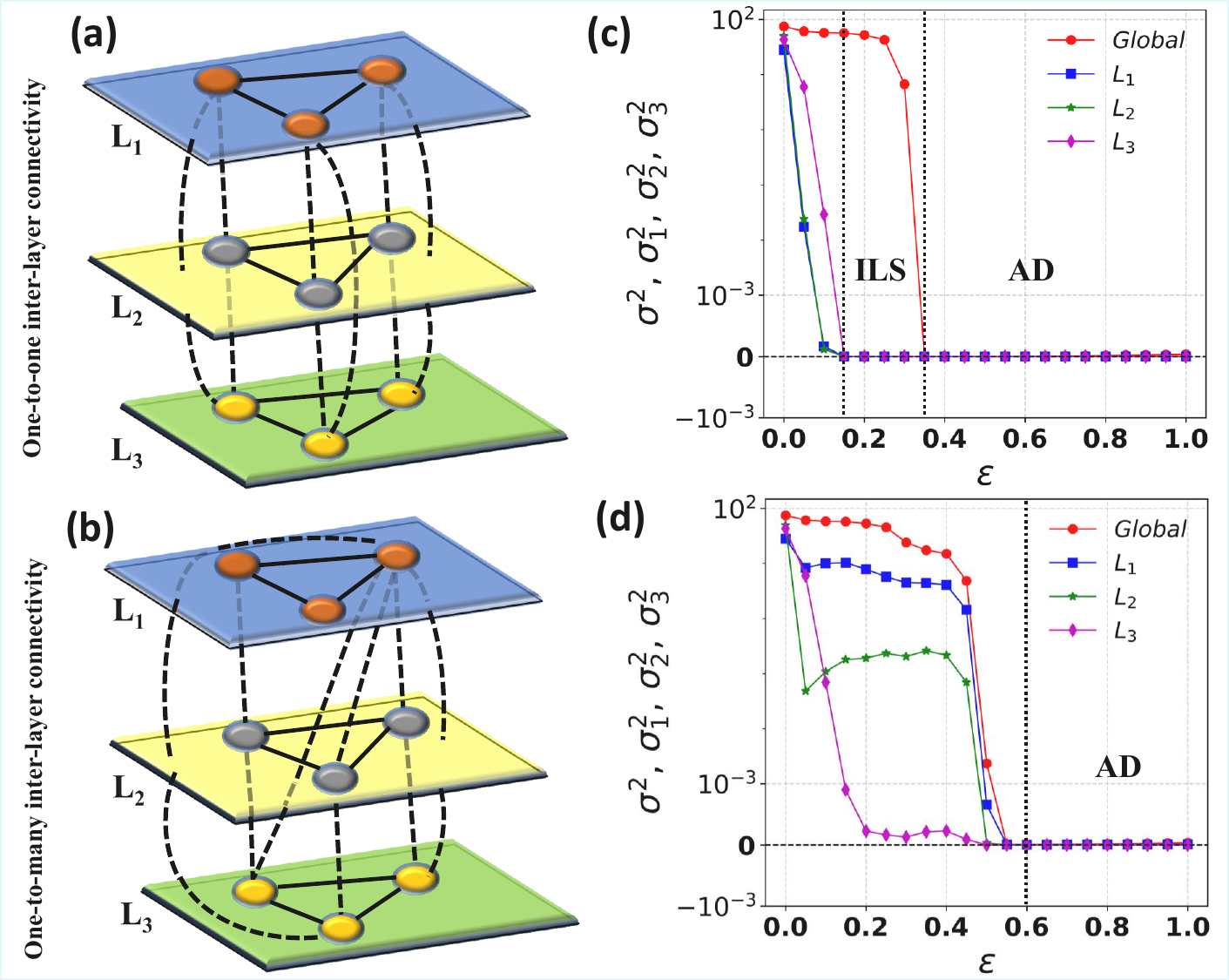}
\caption{{\bf Synchronization for the one-to-one and one-to-many inter-layer connectivity between 3-layers} Sub-figures (c) and (d) show the variation of the synchronization order parameters $\sigma_1, \sigma_2$,  $\sigma_3$, and $\sigma$ for two different inter-layer configurations of $9$ nodes three-layer networks. The schematics (a) and (b) show the connectivity chosen to plot the order parameters. 
For all these plots $N_{1} = N_{2} = N_{3} = 3$, $\omega_1=1.4$, $\omega_2=1.6$, and $\omega_3=0.6 $. All the graphs are plotted for the average of $20$ random initial conditions. 
}
\label{Synchronization-3nodes-3layers}
\end{figure}

\begin{figure} [t]
\includegraphics[width=1\columnwidth]{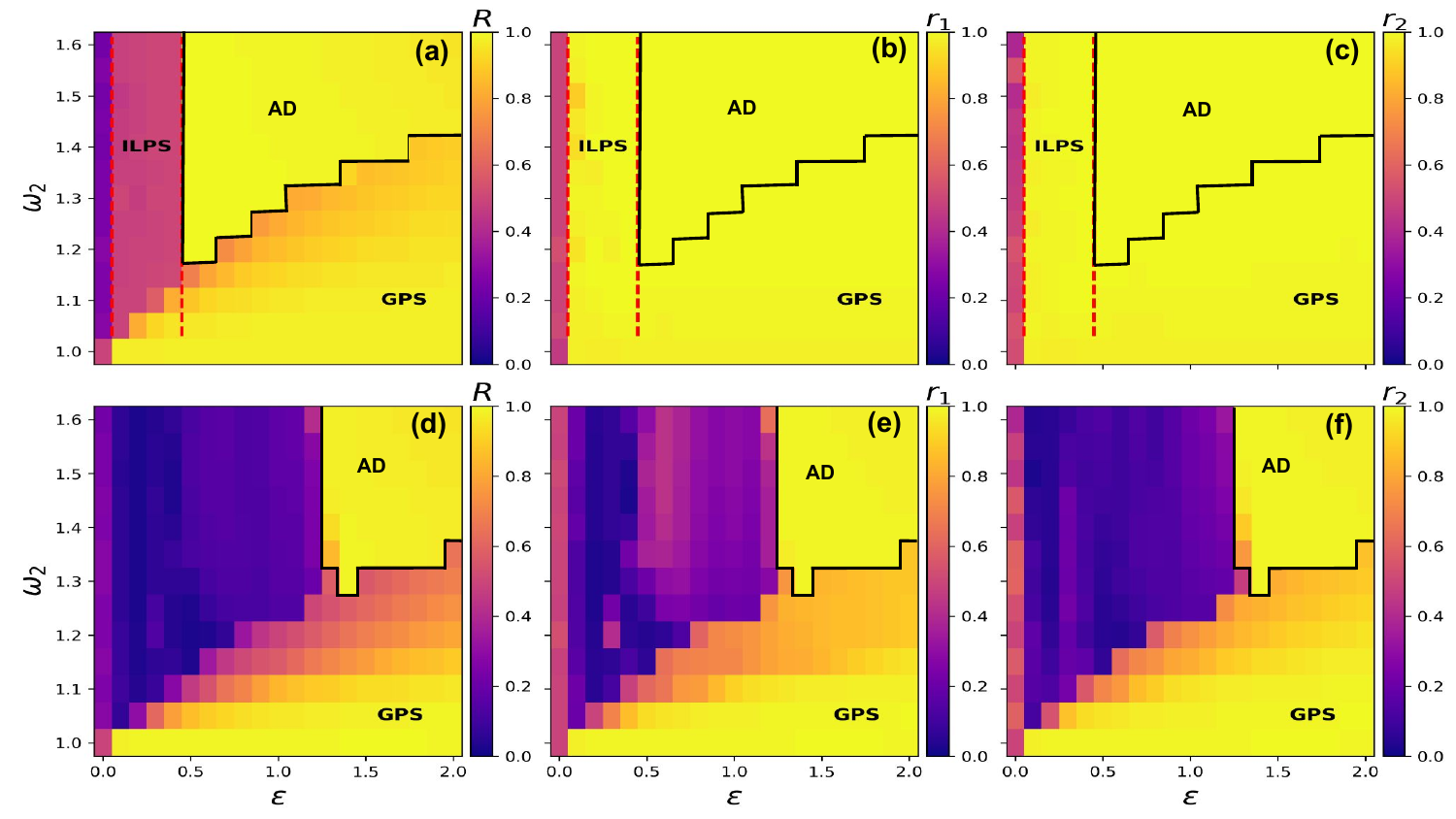}
\caption{{\bf Parameter space for One-to-one and one-to-many inter layer connectivity.} The sub figs show that variation of the  $R$, $r_1$, and $r_2$ versus the $\epsilon$ and intrinsic frequency $\omega_2$ of the second layer in a Bi-layer network of size $N_{1} = N_{2} = 50$. For this plot, we consider both layers to have a ring topology with an average degree $\langle k \rangle = 2$ and $\omega_1 = 1$. For the top (bottom) panel, the inter-layer connectivity is one-to-one (random). All the graphs are plotted for the average of the various different random initial conditions. 
}
\label{Fig_ono-one-color}
\end{figure}

\noindent In this work, we investigate the different emergent dynamics shown by the multilayer networks, considering one-to-one and one-to-many inter-layer connections while maintaining the total number of interconnections constant. By keeping the number of interconnections constant, we can isolate the impact of one-to-one and one-to-many inter-layer connections on the dynamics of each layer. A one-to-one mapping from one layer to another layer (Fig. \ref{Schematic-3nodes}(a)) gives a symmetric inter-layer coupling \cite{inter_intra_type}, whereas the random shuffling of the inter-layer connections, which also causes the existence of one-to-many inter-layer connections, results in asymmetric inter-layer coupling (Fig. \ref{Schematic-3nodes}(b-c)). 
As in real-world networks, the different layers of the multilayer cannot be the same; we consider non-identical layers. The one-to-one inter-layer connections facilitate ILS. This is due to the preservation of intra-layer symmetry, which results from the specific nature of one-to-one connectivity, a characteristic not found in random inter-layer connections.
Additionally, we report another emergent phenomenon, AD, exhibited by non-identical coupled oscillators in different scenarios of inter-layer connectivity and intra-layer topology. Both the one-to-one and one-to-many scenarios exhibit AD; however, AD is achieved at a lower mismatch in parameters and coupling for the one-to-one case compared to the one-to-many case. Moreover, both types of inter-connectivity may exhibit multi-stability in non-identical layers and display remanence, either preserving the exact synchronization of all nodes or periodic phase locking of nodes across different layers, or both. 

\begin{figure} [t]
\includegraphics[width=1.\columnwidth]{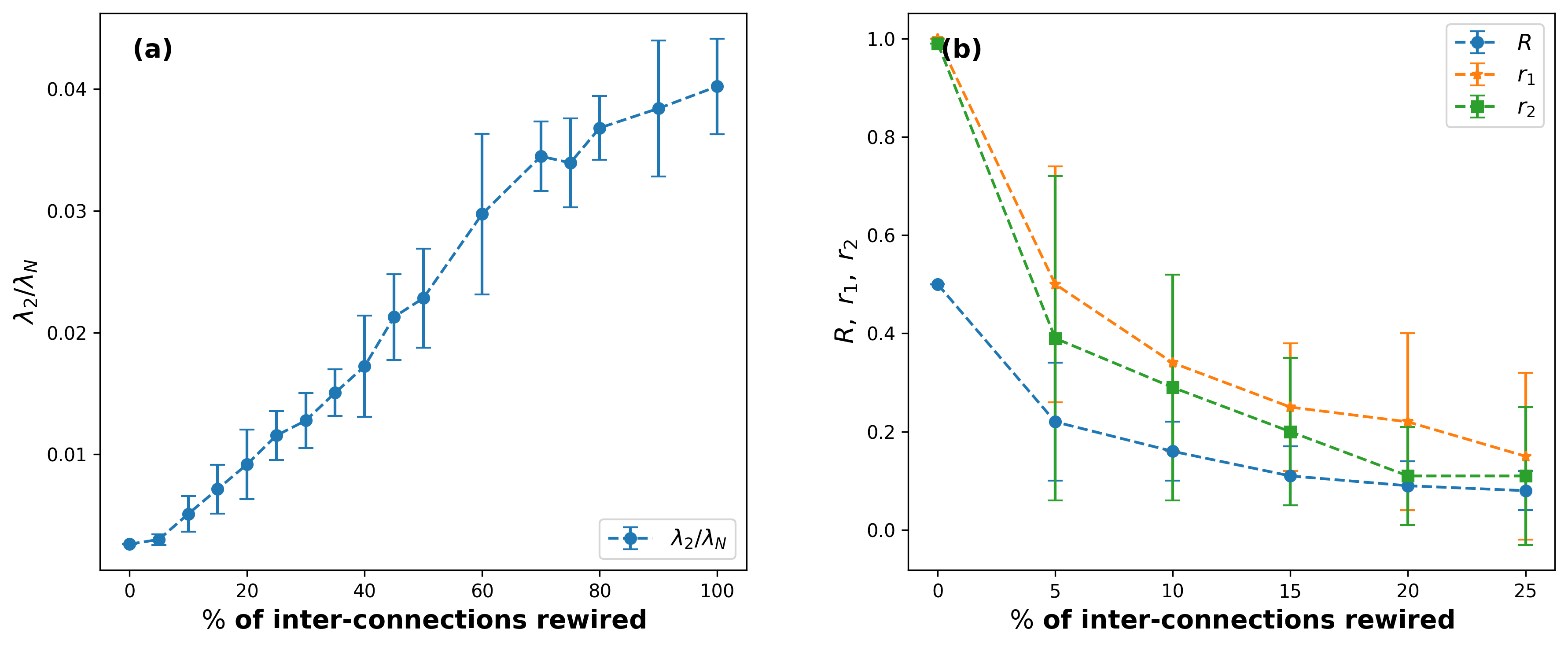}
\caption{{\bf Change in Synchronizability in rewiring from one-to-one to random inter-layer connectivity} The sub figs (a-b) show that variation of the ratio of the first non-zero and the last eigenvalues of the symmetric Laplacian  w.r.t the $\%$ of the connections rewired from the initial stage of the one-to-one mapping. For these plots we consider $N_{l_1} = N_{l_2} = 50$ and ring topology with the average degree $\langle k \rangle =2$ in the both layers. (a) is plotted for the average of the $10$ different random networks, whereas (b) plotted for $5$ different initial conditions for the state. The graphs are plotted for the various network realizations and initial conditions. 
}
\label{Fig_R-lap-percent-rewire}
\end{figure}
\noindent Moreover, as we have explored the coupled dynamics of the R\"{o}ssler oscillators, we also examine the changes introduced by different inter-layer connectivity to the bifurcation. A single R\"{o}ssler oscillator loses stability of the equilibrium via a supercritical Hopf bifurcation, with the emergence of a stable limit cycle, which again loses stability and approaches chaos through a Feigenbaum cascade of period-doubling bifurcations \cite{rasmussen1996bifurcations}. However, the coupled R\"{o}ssler may lose the stability of the fixed points through the torus bifurcation and may show a different route to chaos, such as quasi-periodicity \cite{rasmussen1996bifurcations}. In our study, we find that inter-layer connectivity affects the parameter at which bifurcation occurs; however, it does not affect the type of bifurcations. In both the one-to-one and one-to-many mappings between the different layers, the equilibrium loses stability with increasing connectivity strength via a Hopf bifurcation, which further loses stability via period-doubling bifurcations. Moreover, both cases also exhibit a subcritical Hopf bifurcation. 
\section{Model} \noindent We consider the Multi-layer network having $N_l$ number of nodes in each layer ($l^{th}$). The density and topology can be different in the different layers. We connect the nodes in the different layers with two different schemes; (1) One-to-one, and (2) random. In the One-to-one mapping between the different layers, there is only one connection between the nodes of the different layers. These multilayer networks resemble to the multiplex networks (Fig.\ref{Schematic-3nodes} (a)). In another scheme, the nodes are connected randomly, by keeping the number of the interconnections being equal to the number of the nodes in a layer, and therefore some nodes may receive more than one connection from the other layers (Fig.\ref{Schematic-3nodes} (b-c)) and some may not have any inter-layer connection. The constant number of the connections in the one-to-one mapping and the the random mapping between the different layers allows to see the impact of the only the connection scheme. The networks studied are undirected.  

\noindent Furthermore, to understand the coupled dynamics of the multilayer networks in different inter layer connectivity schemes we use the coupled oscillators model, where the dynamics of the $i$-th oscillator is given as
\begin{equation}
\dot{X}_j = f_1(X_j,\omega_j) + \frac{\varepsilon}{k_j} \sum_{l=1}^N A_{jl}  f_2(X_j, X_l); \; j = 1, \ldots, N
\label{dyn_jth_osc}
\end{equation}
where, $X_j \in R^m$ is $m$ dimensional state vector of the $j$-th oscillator and the parameter $\varepsilon$ defines the strength of overall coupling among the oscillators. $N$ is total number of oscillators in the network. The degree of a node is defined as $ k_{j}$ = $\sum_{l=1}^{N}A_{jl}$, which is the sum of the inter- and intra-layer connections. $f_1: R^m \rightarrow R^m$ provides the dynamics of an isolated oscillator and $f_2$ is the coupling function. We choose $f_1$ as R\"{o}ssler oscillators which are the 3D oscillators capable of giving for rich dynamics. The R\"{o}ssler oscillators are given as follows: $$
\begin{aligned}
\dot{x} &= -\omega y - z, \\[-0.1ex]
\dot{y} &= \omega x + a y, \\[-0.1ex]
\dot{z} &= b + z(x - c)
\end{aligned}
$$
where \(a, b, c\) are system parameters and $\omega$ is the intrinsic frequency of the oscillator. We consider the R\"{o}ssler oscillators in the chaotic regime (\(a = 0.15, b = 0.2, c = 10\)) and the similar intrinsic frequency $\omega$ for the nodes in the same layer but different layers having different different intrinsic frequencies. Moreover, we consider the diffusive y coupling of the R\"{o}ssler oscillators as follows:
\begin{equation}
    f_2(X_j, X_l) = (X_j-X_l), 
\end{equation}
$A$ is the supra-adjacency matrix of the multi-layer-network. For a Bi-layer network defined as follows: 
\[
A = \begin{bmatrix}
A^1 & B^{12} \\
B^{21} & A^2
\end{bmatrix}
\]
where, $A^1$ and $A^2$ are the adjacency matrices for the intra-layer connections, whereas $B^{12}$ and $B^{21}$ is inter-layer connectivity matrix. We consider symmetric intra-layer connectivity, i.e. $B^{12}$ =   $B^{21}$. Moreover, for the one-to-one mapping between the layers $B^{12}$ =   $B^{21}$ = \[
I = \begin{bmatrix}
1 & 0 & \cdots & 0 \\
\vdots &  \ddots & \vdots \\
0 & 0 & \cdots & 1
\end{bmatrix}
\]

\noindent For the random inter-layer connectivity, we rewire the one-to-one connections with the probability $p$ based on the fraction of the connections to be rewired. The random rewiring removes the specificity provided by one-to-one connectivity and creates one-to-many inter-layer connections as well. $A^k_{jl}= 1 = A^k_{lj}$, if oscillators $j$ and $l$ interact in layer $k$, otherwise $A^k_{jl}=0$.  We also consider that the nodes in different layers are non-identical, while considering identical nodes in a layer. Thus, we consider $\omega_l$ as the intrinsic frequency of the nodes in the $l^{th}$ layer.
Further, the phase $\phi$ can be defined as $\phi_j = \arctan\frac{y_j}{x_j} $ and the averaged partial frequency of the $j$-th coupled oscillator is defined as: $\Omega_j(t) = \frac{d \theta_j}{dt}$, which is in general different than the intrinsic frequency $\omega_j$.
\\[1.21ex]
  We evolve Eq.~\ref{dyn_jth_osc} starting from a set of random initial conditions and study the dynamics after an initial transient. We explore the following dynamical states: \\

\noindent {\bf Global synchronized state (GS):} This state corresponds to the exact synchronization between all the nodes and is quantified as follows
\[
\langle \sigma^2 \rangle
= \left\langle \frac{1}{N} \sum_{i=1}^{N} \left( x_i(t) - X(t) \right)^2 \right\rangle_t
\]
where, $\langle \rangle_t$ shows time average. $N$ is the total number of nodes considering all the layers. 
Moreover, the larger sparser networks do not show the exact global synchronization, instead they show the phase locking. 
We say this state as the global phase synchronized state (GPS) and is quantified as follows: 
\begin{equation}
R = \frac{\sum_{i=1}^{N} \sum_{j=1}^{N} \cos(\phi_j - \phi_i)}{N*(N-1)}, 
\end{equation}
where $phi_i$ is the phase of the $i^{\textit{th}}$ oscillator and $N*(N-1)$ is the total number of node pairs in the multilayer network. GPS corresponds to $R\approx1$. 
\\
\\

\noindent {\bf Intra-layer synchronization (ILS):} This state corresponds to the synchronization between the nodes of the different layers irrespective of synchronization between the layers. 
The global exact synchronization in the different layers is demonstrated using the following:
\[
\langle \sigma_l^2 \rangle
= \left\langle \frac{1}{N_l} \sum_{i=1}^{N_l} \left( x_i(t) - X(t) \right)^2 \right\rangle_t
\] where, $N_l$ are the number of the nodes in the $l^{th}$ layer. 
Whereas, the intra-layer phase synchronization (ILPS) is demonstrated by the following:  
\begin{equation}
r_n = \frac{\sum_{i=1}^{N_l} \sum_{j=1}^{N_l} \cos(\phi_j - \phi_i)}{N_l*(N_l-1)}, 
\end{equation}
where, $phi_i$ is the phase of the $i^{\textit{th}}$ oscillator and $N_l*(N_l-1)$ is the total number of node pairs in the $l^{\textit{th}}$ layer. 
ILPS state corresponds to $r_n\approx1$, while R has smaller values.
\\
\\
\noindent {\bf Amplitude death (AD):} This state corresponds to the homogeneous stable steady state at the origin. In the case of the bi-layer networks with the different intrinsic frequencies $\omega_l$, the dynamics for coupled R\"{o}ssler oscillators in a layer ($l$) can be given as follows:
$$
\begin{aligned}
\dot{x_i^l} &= -\omega_l y_i^l - z_i^l, \\[-0.1ex]
\dot{y_i^l} &= \omega_l x_i^l + a y_i^l + \frac{\varepsilon}{k_i^l} \sum_j^N A_{ij} (y_j-y_i^l), \\[-0.1ex]
\dot{z_i^l} &= b + z_i^l(x_i^l - c),
\end{aligned}
$$
where, $k_i^l$ is the total (inter-layer + intra-layer) degree of the $i^{th}$ node of $l^{th}$ layer. For the N nodes in the network, the overall dimension of the system is $3N$. 
The  stability of the fixed points ($\dot{X^*}=0 \forall i \in G$) is done by perturbing the fixed point ($\eta_i = X_i(t)-X^*$). The perturbation  grows as follows: $\dot{\eta_i} = \frac{d(X_i(t)-X^*)}{dt}$.  A Taylor's expansion and linearization gives $\dot{\eta_i} = \eta_i \frac{df(X)}{dX}$ at $X=X^*$. The perturbation will decay and AD will be stable for $\frac{df(X)}{dX}<0$. The matrix presenting $\frac{df(X_1, X_2, .., X_N)}{dX_i}$ in all the dimensions is called as the Jacobian matrix, the sign of the largest eigenvalue of which determines the stability of AD.  For N coupled $3D$ oscillators the Jacobian is  $3N\times3N$ matrix. 

\section{Results} 
\noindent To investigate the differences in dynamical behaviors exhibited by one-to-one and one-to-many inter-layer mappings in multilayer networks, we explore the simplest bi-layer network with all-to-all connections among three nodes in each layer. There are three distinct configuration for this  type of network based on the inter-layer-connectivity: (1) A one-to-one mapping between the nodes of the different layers,  and (2) two one-to-many variants: (i) In the first variant, one node in a layer is connected to two nodes in the other layer, while one node remains without any inter-layer connections, (ii) In the second variant, one node is connected to all nodes in the other layer, and no other nodes have any inter-layer connections (see the schematics in \ref{Schematic-3nodes}). Subsequently, we also present results for larger networks with various network topologies in the layers.
\begin{figure} [t]
\includegraphics[width=1\columnwidth]{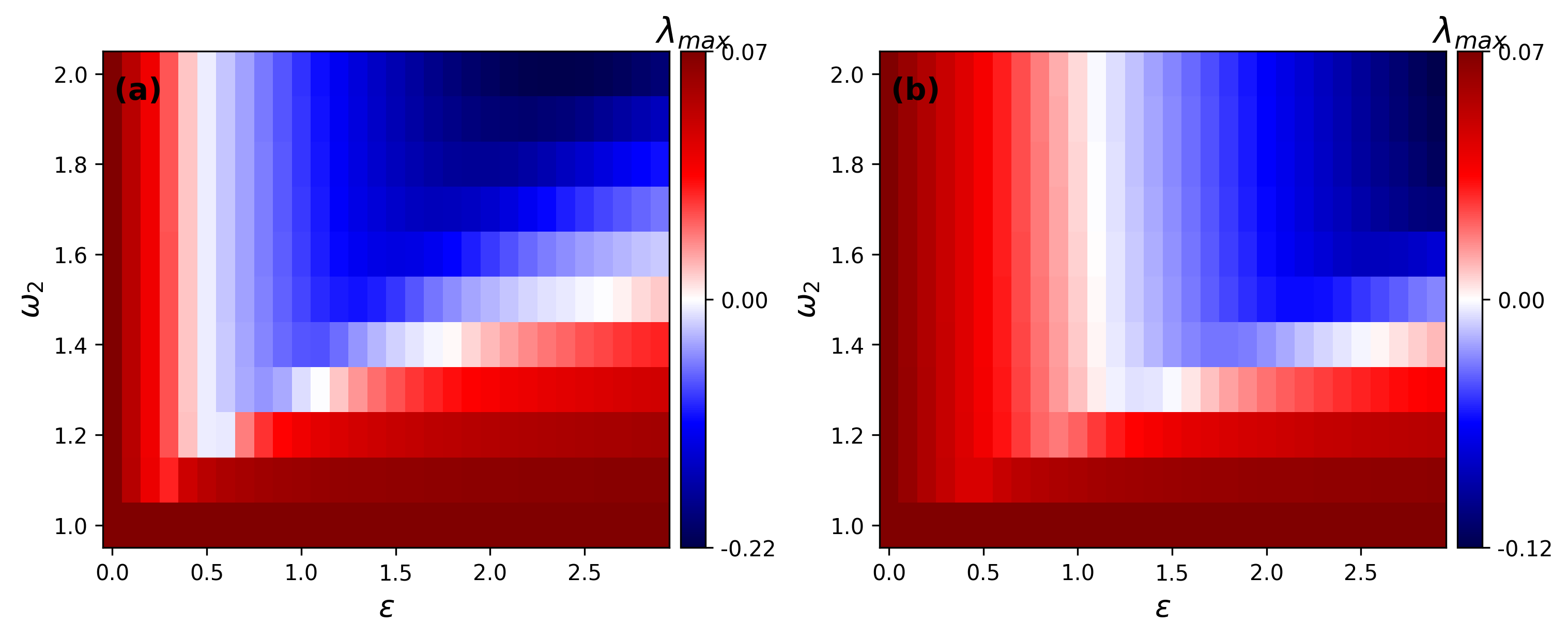}
\caption{
{\bf Parameter space for AD in the Bi-Layer Networks with one-to-one and random inter-layer connectivity} The figure  illustrates the  largest real eigenvalue of the Jacobian matrix at the origin for two senerios: one-to-one (a) and $100\%$ random rewiring of the inter-layer connections, which results random inter-layer topology (b). For the random inter-layer connectivity, the average over $10$ different network configurations is presented. Sub-fig (b) is plotted for average over $10$ network realizations.  }
\label{AD-1d-1d-parameter-space}
\end{figure}
\noindent \subsection{Bifurcations in the ML$\text{s}$ and ML$\text{as}$:} 
\noindent For the simplest $ 3$-node bi-layer networks with different connectivity schemes, we first explore bifurcation with respect to changes in the coupling parameter by studying the largest eigenvalue of the Jacobian at the origin. The Jacobian for the coupled dynamics at the origin is asymmetric; therefore, the eigenvalues are complex. We study the real part of the largest eigenvalue to uncover the stability of the fixed points. We find that the AD range for the three different inter-layer connectivity schemes is different. For the first case of one-to-one connectivity this regime is $0.46\leq \varepsilon \leq 1.78 $, for one one-to-many it is    $0.51\leq \varepsilon \leq 1.94 $, while for the another one-to-many inter-connectivity it is  $0.63\leq \varepsilon \leq 2.33 $ (Figure~\ref{Bifurcation-3nodes} (a)-(c)). In the first case, where all nodes interact with a non-identical node in another layer, the AD occurs at a lower value than in the other two cases, where a lesser number of nodes in one layer interact with distinct nodes in the other layer. Further, as the coupling changes from the mentioned AD range the real part of largest eigenvalue of the Jacobian passes through $0$ and fixed point at origin becomes unstable, the dynamics becomes the small amplitude sinusoidal limit cycle, which oscillates around the origin and the Hopf-bifurcation (HB) occurs irrespective of the inter-connectivity type (Figure~\ref{Bifurcation-3nodes} (d)-(f)). Additionally, stable periodic oscillations emerge after HB, which further lose stability as the coupling increases via period doubling for all inter-layer connectivity. 
Moreover, for all the interconnections, reducing the coupling from the AD state reveals another HB point, leading to unstable periodic oscillations in the one-to-one case  The second case of one-to-many (Fig.\ref{Bifurcation-3nodes}(d, f), which leads to a small window ($\epsilon = 0.4976$ to $0.4718$)  of stable periodic oscillations, emerges for the first one-to-many case (Fig.\ref{Bifurcation-3nodes}(e), which is again followed by the unstable periodic oscillations and the PD bifurcation.  Additionally, in all cases, a Torus point is also observed after the second HB, which is not shown for brevity.

\subsection{Intra-Layer synchronization (ILS) offered by the inter-layer connectivity}  
\noindent Intra-layer synchronization, as mentioned previously, refers to a state in which the nodes within a layer synchronize, while no synchronization is observed between nodes from different layers. To investigate this state, we calculate the synchronization order parameters as mentioned in the methods section. The observation of the $\sigma^2\ne0$ and ($\sigma^2_1 = 0$, $\sigma^2_2 = 0$) shows the existence of this state, which is displayed by figs.~\ref{Synchronization-3nodes}(a)-(c). The one-to-one mapping between the layers gives a wide intra-layer synchronization regime separated by the AD regime. However, in the one-to-many case, the intra-layer synchronization is disturbed, and instead, the synchronized clusters are observed, which are influenced by the symmetry offered by the topology. 
In the case of the one-to-one mapping between the different layers, the $D_3 \times Z_2$ symmetry exists if the nodes of the different layers are identical. The $D_3$ symmetry corresponds to the dihedral group of order $6$. The $Z_2$ symmetry represents the ability to flip the nodes between the two layers. However, because we consider the nodes in the different layers to be non-identical, the $Z_2$ symmetry is lost, leaving only the $D_3$ symmetry intact. 
The broken $Z_2$-symmetry can lead to the creation of certain flow-invariant subspaces where a subset of nodes are synchronized with one another while others are not. These subspaces have been described as poly-diagonal synchronous states~\cite{SGP}. In our case broken $Z_2$ symmetry is responsible for the absence of exact synchronization between the different layers. 
Additionally, in the one-to-many scenario, the $D_3$ symmetry within the layers is disrupted, leading to a breakdown of global-intra-layer synchronization. However, some of the nodes that maintain the symmetry synchronize,  forming clusters. From a graph perspective, the nodes that possess the homophonic relationship (i.e., the exchange of those nodes that preserves the network structure) tend to form clusters. In the one-many case, the two nodes in the second layer that have the same intra-layer neighbor and also receive input from the same node in another layer preserve the symmetry and are synchronized, as shown in Fig.~\ref{Synchronization-3nodes}(e). Similarly, in the one-to-many case 2, where all the nodes of the second layer receive the input from the same node of the first layer, preserve the symmetry and hence synchronize (fig~\ref{Synchronization-3nodes}(f)). In contrast, one node of the first layer that does not preserve intra-layer symmetry due to inter-layer connectivity does not synchronize with the other two nodes in the same layer.  

\noindent Next, we explored the simplest $ 3$-layer networks, having all-to-all connections within the layers and one-to-one and random inter-layer connections \ref{Synchronization-3nodes-3layers}, to demonstrate the existence of the phenomena discussed above in networks with more than $2$ layers. We also find in the $3$-layer networks that the ILS is demonstrated by the layers preserving the homomorphism between all pairs of nodes, regardless of the inter-layer connections. In the three-layer networks with one-to-one interlayer connectivity, ILS is shown for all layers (Fig.~\ref{Synchronization-3nodes-3layers} (a-b)). However, at the same time, the AD is observed at the lower coupling than when some layers have random inter-layer connectivity, giving rise to one-to-many mapping between the layers (Fig.~\ref {Synchronization-3nodes-3layers}(c-d)). This finding further supports the presence of one-to-one connectivity, which allows ILS in all layers but requires a lower coupling strength to exhibit AD, as previously discussed for the bi-layer 6-node networks. It is important to note that, in these three-layer networks, various arrangements of inter-layer connectivity may exist between different layers, and we pick only the case where one-to-one connectivity is present between two layers, while one-to-many connectivity occurs between the other layers.

\begin{figure} [t]
\includegraphics[width=1.\columnwidth]{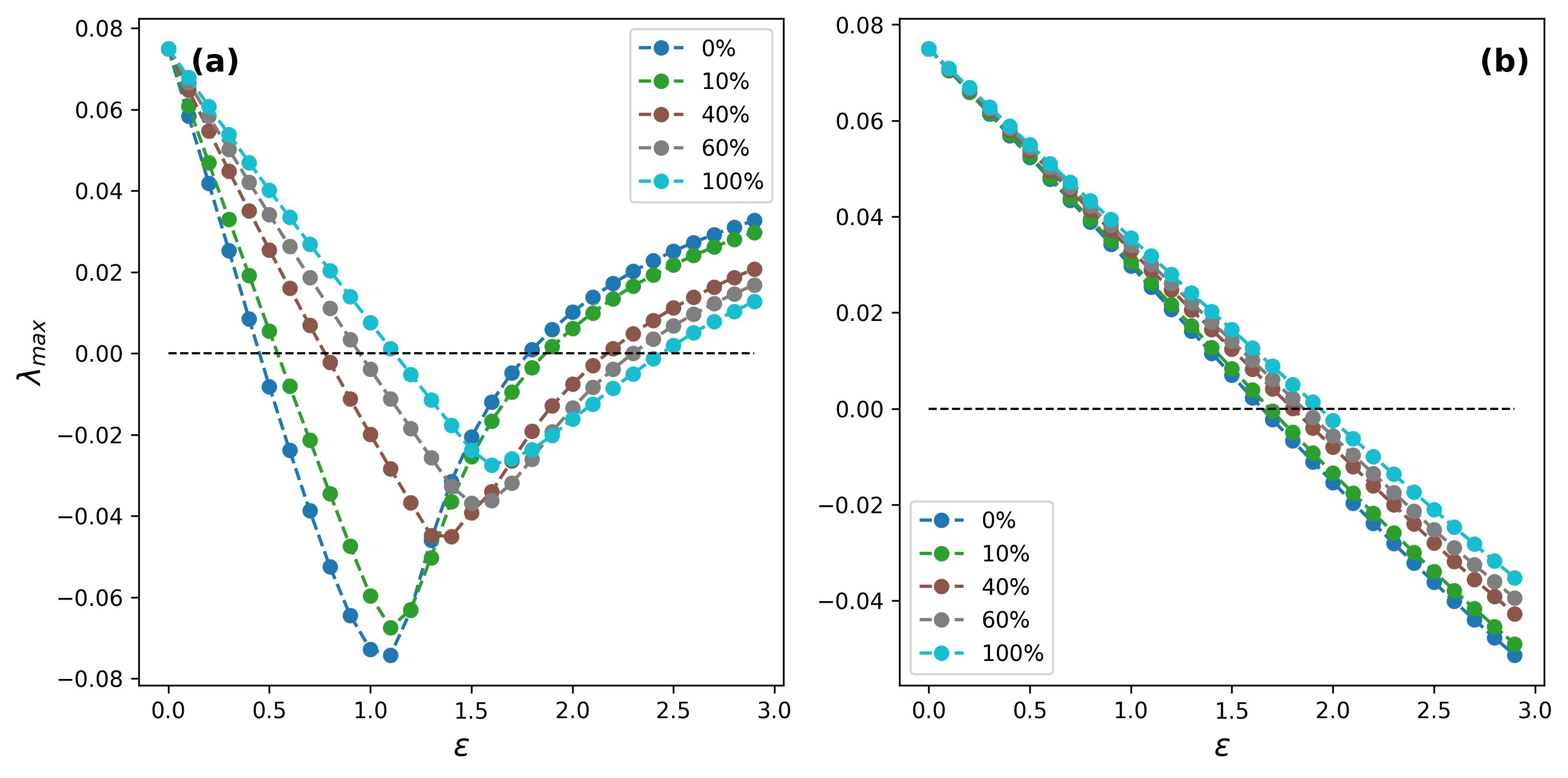}
\caption{{\bf Change in AD regime w.r.t. the rewiring from one-to-one to random inter-layer connectivity.} The sub figs (a-d) show that variation of the largest eigenvalue of the Jacobian for the rewiring percentage, which controls the networks from switching the one-to-one mapping to random inter-layer connectivity. $N_{l_1} = N_{l_2} = 50$ and $\omega_1 = 1$, $\omega_2=1.6$. For this plot, we consider both layers having ring topology with the average degree $\langle k \rangle =2$ in sub-fig(a), $\langle k \rangle =10$ in sub-fig (b). All the graphs are plotted for the $10$ different network realizations. 
}
\label{Fig_AD-rewire}
\end{figure} 
\begin{figure} [t]
\includegraphics[width=1\columnwidth]{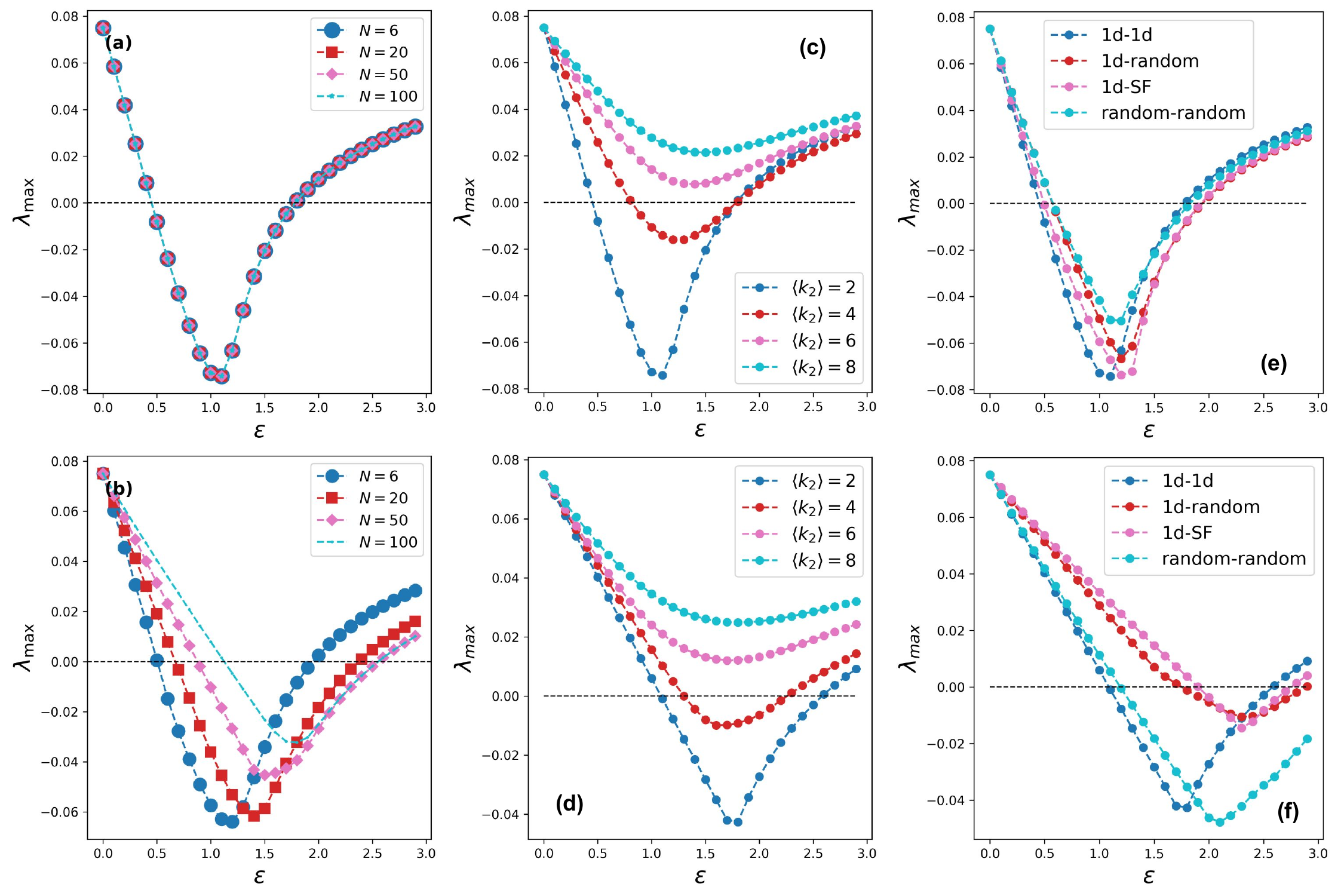}
\caption{{\bf Variation of the largest real eigenvalue of the Jacobian with the different network properties.} Sub-figs.(a-b) demonstrates the change in the size of the network in one-to-one and random inter-layer connectivity, respectively. In this case, we keep the degree of both layers the same $\langle k_1 \rangle =2 = \langle k_2 \rangle$, and both layer has ring topology. (c-d) Demonstrate the change in the density of one of the layer networks in one-to-one and random inter-layer connectivity. In this case, the average degree of the fixed layer is $\langle k_1 \rangle =2$. The topology remains the same, the ring topology in both layers. (c-d) Demonstrate the impact of the topology on AD and consider various cases of networks with one-to-one and random inter-connectivity, respectively. In this case, we fix the density of the layer, $\langle k_1 \rangle =2 = \langle k_2 \rangle$. For all the cases, $N_{1} = N_{2} = 50$.}
\label{jacod-networ-size}
\end{figure}

\noindent Furthermore, we studied the aforementioned phenomena in larger bi-layer networks. As the size of the network increases while maintaining the same density, its synchronizability decreases, and exact synchronization is not observed; however, a weaker form of synchronization, namely phase synchronization, is observed. The phase for the R\"{o}ssler oscillators is defined as mentioned in the model section, and then the order parameters $R$, $r_1$, and $r_3$ are calculated after the initial transient. We plot these parameters for a Bi-layer network having $50$ nodes and average degree $\langle k \rangle =2$ for each layer in a ring topology in Figs.~\ref{Fig_ono-one-color} (a)-(c) in the two-parameter space  $\langle \omega_2 \rangle$ and the coupling strength $\varepsilon$. The frequency of the first layer, $\omega_1$, is fixed to be $1$, so a change in the $\omega_2$ values reflects the change in the difference between the two layers. For these networks, we consider the two cases:(1) the one-to-one mapping between the different layers, and (2) the random-inter-layer connections, while preserving the density of inter-layer connections. 
We start with the layers being identical, where  $\omega_2 = \omega_1 = 1$. In this case, the global phase synchronization (GPS) is observed for both types of inter-layer connectivity, as displayed in the Figs.~\ref{Fig_ono-one-color}(a)-(f). However, when there is a one-to-one mapping between the layers and $\omega_2$ is increased, leading to slightly non-identical layers, we initially observe ILPS at very weak coupling. As coupling strength increases, this switches to the GPS (Fig.~\ref{Fig_ono-one-color}(a)-(c)). Whereas, for the random-inter-layer connectivity, ILPS is never observed ((Fig.~\ref{Fig_ono-one-color}(d)-(f))) even at the lower coupling values.  Instead, GPS is observed, which depends on the frequency mismatch between the different layers and the strength of the connectivity. At the higher coupling values, for the larger mismatch between the layers, in both the type of inter-layer connectivity, instead of ILPS or GPS, the AD state emerges (Fig.~\ref{Fig_ono-one-color}). We will discuss the AD state in detail in the coming section. 
\begin{figure} [t] 
\includegraphics[width=0.99\columnwidth]{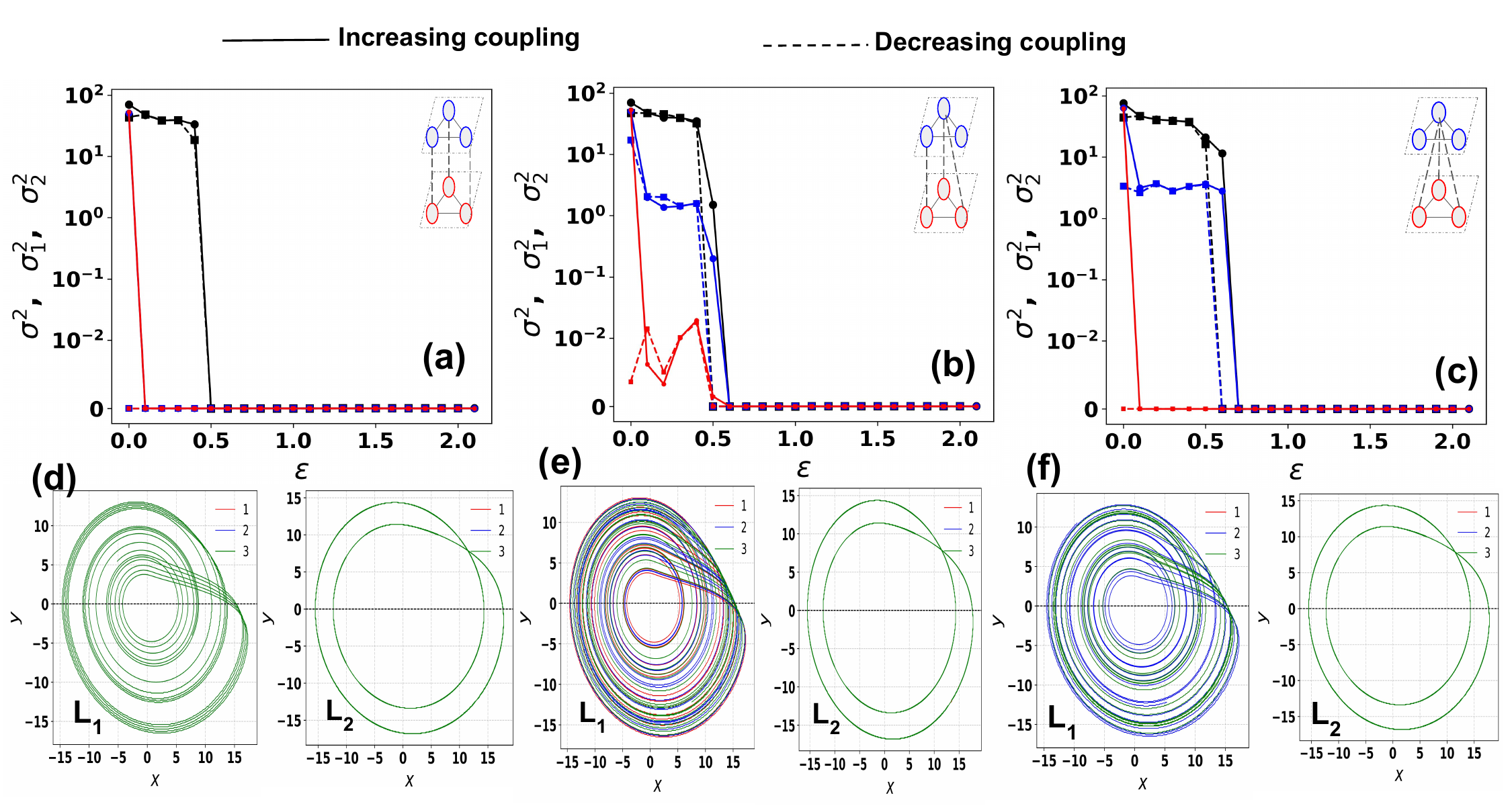}
\caption{({\bf Hysteresis plot for the multilayer networks with different inter-layer connectivity} The sub-figures (a-c) demonstrate the variation of $\sigma$ (black squares), $\sigma_1$ (red squares), and $\sigma_2$ (blue squares) w.r.t. $\varepsilon$ in the forward (solid lines) and backward (dashed lines) direction. The sub-figures (d-f) show the trajectories of the oscillators in the x and y planes. $L_1$ and $L_2$ represent layer 1 (blue circles in the schematic) and layer 2 (red circles in the schematic), respectively. For these plots, we kept $\omega_1 = 1$ and $\omega_2 = 1.6$.     }
\label{Fig_Hyst1}
\end{figure}

\noindent Furthermore, to examine how a gradual change in interconnectivity, from one-to-one to random, affects synchronization, we plot the ratio of the first non-zero and last eigenvalues ($\lambda_2/\lambda_N$) of the normalized Laplacian ($L_{ij} = I-D^{-1/2}AD^{1/2}$). This ratio has been found to be associated with the synchronizability of the networks \cite{Laplacian-eigen-synchronizability}. The networks with the higher $\lambda_2/\lambda_N$ values are better synchronizable. 
We find that a systematic rewiring of the inter-layer connections from one-to-one to a random pattern, while both layers maintain their lattice structure, leads to an increase in this ratio (fig.~\ref{Fig_R-lap-percent-rewire} (a)). This indicates that networks with random inter-layer connections are better synchronizable than those with one-to-one inter-layer connections. However, when the nodes in the different layers are non-identical at the lower couplings instead of an increment, a decrement in the order parameter values ($R$, $r_1$, and $r_2$) is observed, with the $100\%$ rewiring completely suppressing the intra-layer phase synchronization exhibited by the bi-layer networks with the one-to-one mapping (fig.~\ref{Fig_R-lap-percent-rewire} (b)). This suggests that, in the case of non-identical layers, graph homomorphisms are better estimators of synchronizability than the Laplacian eigenvalues. 
\noindent \subsection{AD in the multilayer networks with different inter-connectivity}
\noindent As mentioned in the model section, we study the AD state by studying the eigenvalues of the Jacobian at the origin. 
We find that the AD regime depends on the inter-layer connectivity; a one-to-one inter-layer connectivity gives AD for the lesser mismatch in the layers and at lower coupling than the random inter-layer connectivity. For example, as plotted in Figs. \ref{AD-1d-1d-parameter-space}, in a network with the ring topology and $50$ nodes in each layer, with one-to-one connectivity,  AD starts happening for a mismatch in the intrinsic frequencies of $0.2$ (Figs. \ref{AD-1d-1d-parameter-space}(a)), whereas, for the random inter-layer connectivity, the AD starts for a higher mismatch of $0.3$ (Figs. \ref{AD-1d-1d-parameter-space}(b)). Additionally, the strength of coupling required for AD is lower for one-to-one connectivity than for random inter-layer connectivity, which also includes one-to-many connections. For the mentioned mismatch, it begins with $\varepsilon \approx 0.5$ and $\varepsilon = 1.2$ for one-to-one and random inter-layer connectivity, respectively. A higher frequency mismatch always leads to AD irrespective of the inter-layer connectivity. 
Moreover, a systematic random rewiring process, starting from one-to-one connectivity and gradually shifting to $100\%$ random rewiring, gradually shifts the AD range to higher coupling values. For example for the sparse networks ($\langle k \rangle =2$) with ring topology, the $\varepsilon$ value shifts from $\approx 0.5$ to $\approx 1.2$ as the inter-connectivity shifts from one-to-one to $100\%$ random rewiring ((Fig.\ref{Fig_AD-rewire} (a)). This effect is also observed in denser networks; however, the shift in the coupling range is less pronounced. For instance, in the ring topology with an average degree of  $\langle k \rangle =10$ in each layer, a frequency mismatch of $0.4$ results in the AD starting at $\varepsilon \approx 1.7$ for one-to-one coupling. This value shifts to $2$ with $100\%$ rewiring (Fig.\ref{Fig_AD-rewire} (b)). 

\begin{figure} [t] 
\includegraphics[width=0.5\columnwidth]{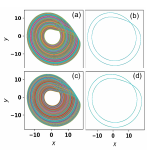}
\caption{{\bf Time-series for 1d-SF }  For this plot, we consider a bi-layer network having ring topology in the first layer and scale-free topology in the second layer, with the average degree in both layers $\langle k \rangle =2$. (a, b) corresponds to the one-to-one, and (c, d) is for the one-to-many inter-layer connectivity. $N_{l_1} = N_{l_2} = 50$. $\omega_1 = 1$ and $\omega_2 = 1.6$.      
}
\label{Fig_Hyst2}
\end{figure}

\noindent Furthermore, we investigate how the size of the networks affects the AD range and find that for one-to-one connectivity, this range remains invariant of the size of the network (Fig.~\ref{jacod-networ-size} (a)). In contrast, for random inter-layer connections, the variation depends on the network size and range shifts to a higher coupling value as the size increases (Fig.~\ref{jacod-networ-size} (b)). Moreover, irrespective of the inter-layer connectivity, a change in density of only one layer also shifts the $\varepsilon$ range for AD to higher values (Fig.~\ref{jacod-networ-size} (c, d)). A coupling of a sparse layer with very dense networks can also mitigate the AD at the lower mismatch and coupling values. For example, a sparse layer with the average degree $\langle k \rangle =2$ and ring topology, when coupled with another layer with ring topology and $\langle k \rangle >4$, the AD regime observed for a lower density match disappears (Fig.~\ref{jacod-networ-size} (c, d)) for both the one-to-one and random inter-layer connectivity. 
Moreover, we also studied the impact of the topology of the different layers on the AD regime. For this, we studied Bi-layer networks, where one layer is a lattice and the other is either random or scale-free. Additionally, we examined the case where both layers have a random topology. We generate random networks using the E-R method and scale-free networks using the B-A algorithm. We find that a change in the topology of the layers affects the AD regime. For one-to-one mapping, this effect is not as pronounced as for the random inter-layer connectivity (Fig.~\ref{jacod-networ-size} (e, f)). For random inter-layer connectivity, the AD range is significantly different and starts at the lower coupling values for the lattice-lattice case (Fig.~\ref{jacod-networ-size} (f)). The random topology in both layers leads to the widest regime for AD. In the case of scale-free networks, the AD regime for random inter-layer connectivity is susceptible to whether the hub nodes receive input from the other non-identical layers. When most of the hubs receive input from the other layer, the AD regime shifts to lower values. For Fig.~\ref{jacod-networ-size} (f), we plotted an average over all such possible networks.

\subsection{Hysteresis and remanent order in different inter-layer connectivity} 
\noindent 
To observe the hysteresis in the coupled dynamics of the multilayer network, we start with $\varepsilon =0$, and evolve the dynamics by increasing the coupling up to a value $2.1$, and then decrease the coupling values gradually to $\varepsilon = 0$ to investigate if the system traces the previous path or not. Here,  instead of the random initial condition for the states of the oscillators as considered for studying the synchronization phenomenon as discussed in the previous section, the initial state for the subsequent coupling is considered as the state at the previous coupling value. This is analogous to the slow heating (increase in coupling) and cooling (decrease in coupling), and is preferred to avoid the sudden perturbations to the fixed states by the random state variables. We study this phenomenon for the small $6$-node multilayer networks, as discussed previously, and for larger networks with various topologies.  The networks of small sizes show a monotonic decrease in the global order parameter ($\sigma^2$) as the coupling increases; however, $\sigma_1^2$ and $\sigma_2^2$ drop to the $0$ value as the coupling becomes $0.1$. In contrast, $\sigma^2$ remains non-zero (Fig.~\ref{Fig_Hyst1} (a), showing the emergence of the intra-layer synchronization, as discussed previously. $\sigma^2$ vanishes for a higher coupling value; however, this state is the AD state. Next, we decrease the coupling from a $2.1$ value and observe that, although the global order parameter traces the same path as it did for the increase in coupling, $\sigma_1^2$ and $\sigma_2^2$ exhibit different behaviors. Instead of becoming non-zero at 
$\varepsilon=0$, they remain zero. Displaying that even when there is no coupling, the oscillators exhibit synchronized dynamics in both layers. We further explore the timeseries of this state and find it to be periodic synchronized for the faster layer (Fig.~\ref{Fig_Hyst1} (d)).

\noindent Next, we change the inter-layer connectivity from one-to-one to random, which alters the path of the order parameters, as discussed above. We observe a switch from the intra-layer synchronization state to AD at the higher coupling values. However, when coupling is reduced, the transition from AD back to the intra-layer synchronization occurs at the same coupling value as exhibited by the one-one case, indicating a hysteresis effect in the synchronization parameter $\sigma$ ( see Fig.~\ref{Fig_Hyst1} (b)). Additionally, the global remnant synchrony seen in the one-to-one case is lost; instead, cluster synchronization between nodes with a homomorphic relationship can be observed. Nonetheless, the faster layer continues to exhibit periodic dynamics (see Fig.~\ref{Fig_Hyst1}(e)), which again show cluster synchrony, rather than global synchrony. 

\noindent In the one-to-many case (2), as sketched in Fig.\ref{Schematic-3nodes}(c), we observe a mixed behavior that combines the characteristics of the previously discussed two cases. Hysteresis in the global order parameter is present, and the layer maintaining the $D_3$ symmetry, due to its inter-connectivity with the same node across different layers, shows synchronized remanence. At the same time, another layer loses it (see Fig.~\ref{Fig_Hyst1} (c)). Additionally, the faster layer, which also exhibits synchronized dynamics, shows periodic behavior, as observed in the previous two cases (Fig.~\ref{Fig_Hyst1} (f)). 

\noindent Furthermore, we studied the Hysteresis for larger networks and observed similar trends to those shown by the small network discussed above. However, instead of Hysteresis in the global order parameter, it is now observed in the order parameter for phase synchronization. In the one-to-one connectivity, the slower layer exhibits remanence in the phase-synchronized oscillators. In contrast, the fast layer displays phase-synchronized periodic oscillations, regardless of the topology of the different layers. 
Additionally, in the one-to-many connectivity, Hysteresis is observed; however, remanence is lost in the slower layer, while the remanence of the periodic oscillation in the faster layer is retained. For example, Fig.~\ref {Fig_Hyst2} displays the dynamics of the oscillators in a bi-layer network with both lattice and scale-free topologies in the different layers. It shows that when the coupling becomes zero after the adiabatic cooling, the one-to-one connectivity gives the phase synchronized oscillations for the slower layer (Fig.~\ref{Fig_Hyst2} (a)). In contrast, faster layers exhibit the periodic phase synchronized periodic oscillations (Fig.~\ref{Fig_Hyst2} (b)), as also discussed for the small network previously. However, the phase synchronization vanishes for the random inter-layer connectivity, but for the faster layer, the periodic dynamics are still displayed (Fig.~\ref{Fig_Hyst2} (c), (d)). This periodic state may represent splay state (phase locked incoherent state) rather than a  phase-synchronized state, as the order parameter does not display higher values.

\section{Discussion and conclusion} 
\noindent The symmetries of a network have previously been shown to support cluster synchronization and global synchronization in random multilayer networks \cite{Symmetries_syn-multilayer}. In our study, we explored in detail the different dynamics exhibited by the various inter-connectivity in multilayer networks with non-identical layers. We find that the $D_N$ symmetry offered by one-to-one connectivity enables intra-layer synchronization. In contrast, the random inter-layer connections destroy the $D_N$ symmetry of the Layers, and thus the ILS is not observed in multilayer networks with random inter-layer connections. Additionally, the random inter-layer exhibits a broader range of parameters for which it does not display synchronization, yet it still shows oscillations. Whereas, in the one-to-one case, the incoherent oscillation range is limited, the AD range is wider, showing a more drastic impact on oscillation quenching due to interaction with non-identical nodes in another layer. 
The reason for this could be that, for one-to-one mapping, all nodes in a layer interact with a non-identical node from another layer. In contrast, in the case of random inter-layer connectivity, some nodes do not interact with nodes in another layer. However, some nodes interact with more than one node from the other layer. The oscillating nodes that do not directly interact with the non-identical nodes retain oscillation for lower couplings. Their interaction with the nodes receiving input from the non-identical nodes may prevent oscillation quenching at lower couplings and parameter mismatches.  

\noindent Moreover, the parameter regime for AD in a one-to-one inter-layer connectivity regime is independent of the network size. It mainly depends on the density and topology of the different layers. AD is not desirable for some systems that rely on the rhythms of its different units, such as the brain. In such a system, a one-to-one mapping between the layers should be avoided to mitigate AD, as this inter-layer connectivity minimizes the mismatch and couplings required for AD. Future research could focus on exploring such a network, along with utilizing models specific to this system for further investigation. However, in engineering systems where the AD is required to suppress noise or control the system, a one-to-one mapping can be considered. Although AD has previously been reported for networks with mismatched parameters \cite{AD_strogatz, Chaos-supression}, our work adds to it by commenting on the role of inter-layer connectivity in a multilayer framework on the feasibility of AD. 

\noindent Furthermore, we demonstrate that inter-connectivity leads to multi-stability in the non-identical multilayer networks. The coupled non-identical R\"{o}ssler oscillators have previously been shown to exhibit the dynamics hysteresis \cite{Rossler-dynamic-hysteresis}, however, not in the context of the multilayer networks. 
We find that one-to-one inter-connectivity can give rise to permanent memory in all interacting layers, with the faster layer exhibiting remanence in non-synchronized periodic dynamics of chaotic oscillators, and the slower layer showing remanence in synchronized oscillations. Although the random inter-layer connectivity also exhibits Hysteresis, it is not as pronounced as for the one-to-one connectivity, and the synchronized remanence shown by the slower layer is not observed in this case. In contrast, the remanence shown in the periodic dynamics of the faster chaotic oscillators is retained. Hysteresis is one of the important phenomena shown by complex systems. The origin of the Hysteresis lies in the existence of multiple path-dependent metastable states. Remanence, the persistence of memory in the system, can also be reflected in the system exhibiting Hysteresis. The existence of memory in coupled systems shows that not only adaptation, but also multi-stability, can be responsible for permanent memory in real-world networks. In the context of the brain, our results are particularly significant. As they suggest, when a group of neurons interacts with other systems, which could also be an external stimulus, this interaction can lead to the formation of lasting memories associated with that stimulus, based on their inter-connectivity. Additionally, it could serve as one of the mechanisms for achieving Hysteresis and memory in technical systems.

\noindent \section*{Acknowledgment} 
\noindent AS and SD thank the Department of Science and Technology (DST), Government of India, for their financial support through the DST INSPIRE Faculty grant (IFA-21-PH-276). AS, AR, SD and DP are also thank IISER Tirupati for providing the necessary infrastructure and the High-Performance Computing (HPC) facility. AS acknowledges Dr. Samir Sahoo from San-Diego State University for useful discussions.  A. Palacios was supported by MURI ONR (Grant No. N000142412547)

\bibliographystyle{unsrt}  
\bibliography{Refrence_one_one_work}  
\end{document}